\documentclass[aps,pra,eqsecnum,twocolumn,showpacs,superscriptaddress]{revtex4}
\usepackage[]{graphicx}
\usepackage{amsmath}
\usepackage{color}
\begin{document}
\title{Identifying quantumness via addition-then-subtraction operation}

\author{Su-Yong Lee}
\affiliation{Centre for Quantum Technologies, National University of Singapore, 3 Science Drive 2, 117543 Singapore, Singapore}

\author{Changsuk Noh}
\affiliation{Centre for Quantum Technologies, National University of Singapore, 3 Science Drive 2, 117543 Singapore, Singapore}

\author{Dagomir Kaszlikowski}
\affiliation{Centre for Quantum Technologies, National University of Singapore, 3 Science Drive 2, 117543 Singapore, Singapore}
\affiliation{Department of Physics, National University of Singapore, Singapore}

\date{\today}

\begin{abstract}
We propose a measure of quantumness based on an addition-then-subtraction operation. We demonstrate how this measure can distinguish between classical and bosonic particles by investigating in detail multi-particle bosonic systems. Experimental schemes implementing this measure for bosons in all-optical and atom-cavity systems are provided. We also apply this measure to single-mode fermionic systems. 
\end{abstract}

\maketitle

\section{Introduction}
Quantum particles exhibit wave-particle duality \cite{Scully}. The wave nature allows quantum particles not only to show
interference in a double-slit experiment but also to be indistinguishable from each other if they are in the same energy state.
Quantum particles are classified into bosons and fermions showing bunching and anti-bunching effects, respectively.
A large number of noninteracting bosons can occupy the lowest energy state, while only one fermion can occupy a particular energy state due to the Pauli exclusion principle.
Quantum particles exhibit entanglement \cite{Einstein}, which is an essential resource in quantum information processing \cite{Nielsen}.
For example, by injecting single photons (bosons) to the input modes of a $50:50$ beam splitter, one can generate a well-known entangled state $|2,0\rangle-|0,2\rangle$ by Hong-Ou-Mandel
interference \cite{Hong}. For fermions, by injecting two electrons to two mesoscopic electron beam splitters \cite{Liu}, 
i.e., $\widehat{B_M}_{bc}\widehat{B_M}_{ab}|1\rangle_a|0\rangle_b|1\rangle_c$, one can generate a Dicke state of three qubits with two excitations \cite{Dicke,Tomek}, $\frac{1}{\sqrt{3}}(|1,1,0\rangle+|1,0,1\rangle+|0,1,1\rangle)$.
On the other hand, classical particles are distinguishable even in the same energy state, resulting in different statistics. 

The differences among classical, bosonic, and fermionic particles have been investigated with the statistical behavior of the particles at a lossless beam-splitter, for example \cite{Loudon}.
When each particle is injected into a $50:50$ beam splitter, the output modes generate different statistics, i.e., classical [$P(1,1)=1/2,~P(2,0)=P(0,2)=1/4$], Bose [$P(2,0)=P(0,2)=1/2$], and Fermi [$P(1,1)=1$] statistics.
Other works investigating the difference between these three different behaviors include $(1)$ measuring quantumness via an anticommutator \cite{Fazio}, where the quantumness of any two quantum states was quantified by the nonpositivity of the anticommutator, and $(2)$ measuring bosonic and fermionic properties with the difference of vacuum state probability via addition-then-subtraction operation \cite{Pawel}, where the measuring properties were restricted to vacuum and single-particle probabilities.
For bosons, there have been investigations on nonclassical properties by photon
addition-then-subtraction operation $\hat{a}\hat{a}^{\dag}$ \cite{Kim, Yang, Suyong}.
Moreover, the sequential operations of addition and subtraction are useful in probing quantum commutation rules \cite{Parigi,KJZPB,ZPKJB},
improving entanglement properties in continuous variable systems \cite{Li, Suyong1}, enhancing the degree of nonlocality \cite{Park},
achieving noiseless amplifier \cite{Zavatta}, 
and quantifying bosonic (fermionic) behavior in composite particle systems \cite{Pawel}.

Here we propose a scheme to compare the three types of particles in a single mode via addition and subtraction operations and suggest possible experimental implementations.
Suppose one is given a device that adds and subtracts a given type of particle. This device works very differently for the three types of particles.
For classical particles, the addition and subtraction operations work deterministically, whereas 
for quantum particles, the operations can only be carried out probabilistically, e.g., addition and subtraction of photons \cite{Kim}.
Thus, using an addition-then-subtraction operation, we can distinguish the quantum particles from the classical particles and quantify 
quantumness through their indistinguishability. 
 We define the degree of indistinguishability as quantumness via addition-then-subtraction operation. It represents the tendency of particles to stay together in a mode. Classical particles are distinguishable so that we can tell them apart. The degree of indistinguishability for classical particles is fixed, regardless of the number of particles.  In contrast, bosonic particles are indistinguishable, so that we cannot tell them apart.  Thus the degree of indistinguishability for bosonic particles increases with the number of particles. Fermionic particles obey the Pauli exclusion principle so that each particle should occupy one mode. Due to this limit in the occupation number, the degree of indistinguishability for fermionic particles is bounded. 

 We focus on multiparticle probabilities which are useful in quantifying the degree of indistinguishability for bosonic particles.
For photons,  the degree of indistinguishability (upon adding and then subtracting a photon) increases with the increasing mean photon number and therefore the deviation from classical behavior increases. 
This feature is also present in the case of coherent states in spite of the fact that their fractional uncertainty in photon number and their phase uncertainty decrease, which is considered to be a classical feature of this class of states. This shows us that one has to be careful of what to measure to detect quantumness of bosonic systems. Restriction to measurements of a photon number distribution fails to capture a highly nonclassical behavior that stems from indistinguishability of bosons.

This paper is organized as follows. Section II begins with an identification of classical and quantum particles from the particle viewpoint and provides a generalized operator description.
In Sec. III, we propose a measure to test bosonic and fermionic properties via an addition-then-subtraction operation.
Furthermore, we suggest implementable schemes in an atom-cavity system as well as in an all-optical system.
We conclude in Sec. IV.

\section{Identification of classical and quantum particles}
We consider addition and subtraction operations to distinguish quantum particles from classical ones.
In classical mechanics, addition and subtraction operations are independent of the initial state, such that
the initial state does not change after the sequential operation, i.e., the addition-then-subtraction or subtraction-then-addition operation.
In quantum mechanics, addition and subtraction of a boson (photon) depends on the initial state. For example, when there are $n$ bosons initially, addition operation is defined as $\hat{a}^{\dag}|n\rangle=\sqrt{n+1}|n+1\rangle$, where $\sqrt{n+1}$ indicates the probability amplitude that we can identify the photon we added in the final state after addition.
Subtraction operation is defined as $\hat{a}|n\rangle=\sqrt{n}|n-1\rangle$, where $\sqrt{n}$ indicates the probability amplitude that we can identify the photon we subtracted in the initial state before subtraction. 
After the addition-then-subtraction operation ($\hat{a}\hat{a}^{\dag}$) or subtraction-then-addition operation ($\hat{a}^{\dag}\hat{a}$) on a state $|n\rangle$, 
the corresponding particle state becomes $(n+1)|n\rangle$ or $n|n\rangle$, 
where the prefactors ($n+1$) and $n$ exhibit the tendency for bosons to stay together in a mode.

Differences between classical and quantum particles are clearly seen from the effects of addition and subtraction of particles on a single-particle state.
Note that one has to start with at least one particle because the process of one-particle subtraction is not possible when there is no particle to begin with.
For one-particle \it{subtraction then addition}\rm, there is no difference between classical and quantum particles if we start with one particle: The initial particle is merely taken away from the initial place and then returned to the same place.
For one-particle \it{addition then subtraction}\rm~ from one particle, however, there is a difference between classical and quantum particles.
Of course, a classical particle returns to the initial state after the operations, but this is not the case for a quantum particle.
After adding one quantum particle to the initial one, it is impossible to deterministically subtract the same particle since quantum particles are indistinguishable. The situation is depicted in Fig.~1, 
where we associate the indistinguishability with the overlap between the quantum particles.
\begin{figure}
\centerline{\scalebox{0.35}{\includegraphics[angle=270]{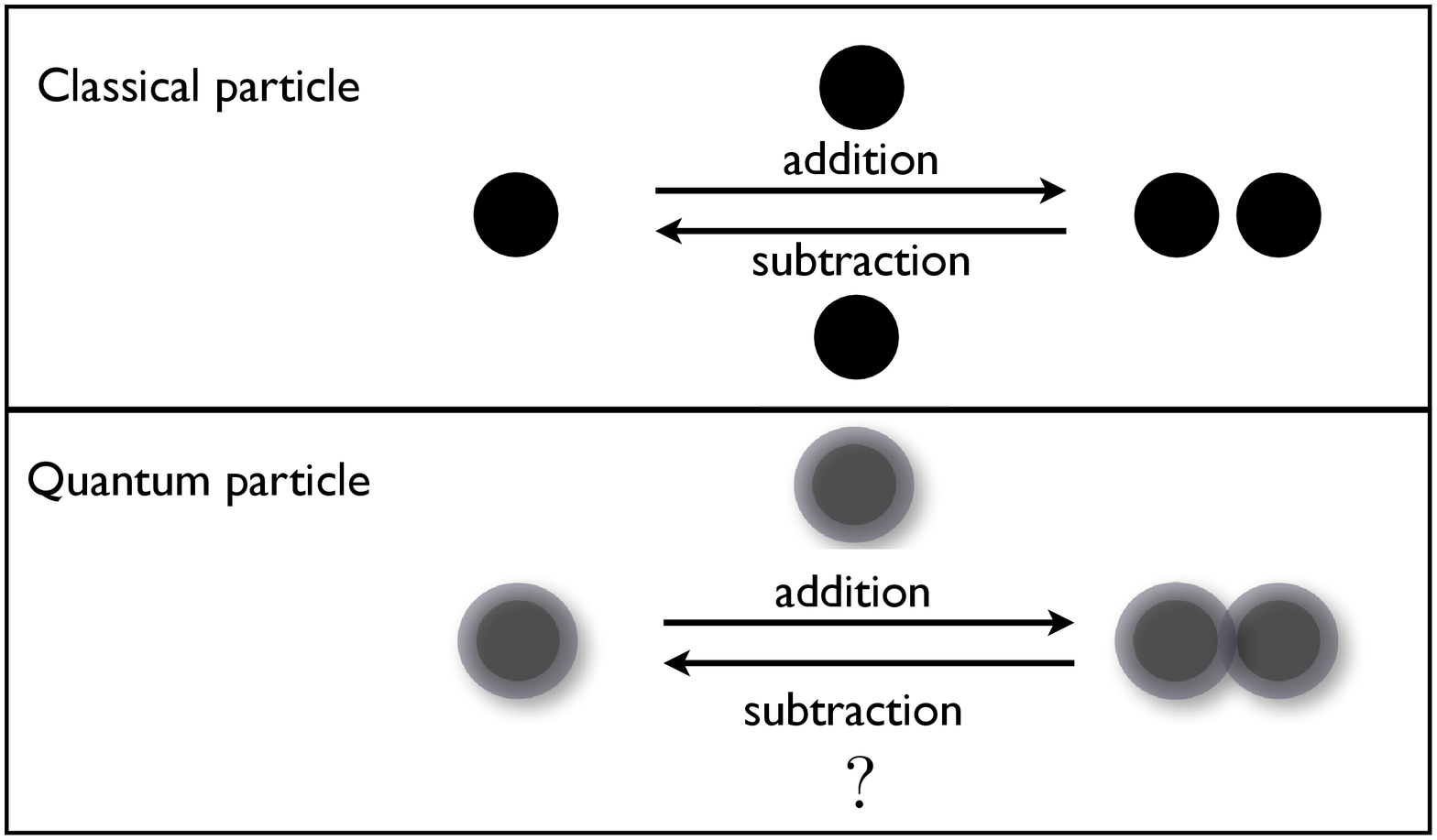}}}
\vspace{-0.8in}
\caption{Identification of classical and quantum particles via addition then subtraction. Assume that initially we have two independent particles.
Classical particles are distinguishable and quantum particles are indistinguishable. }
\label{fig:fig1}
\end{figure}

A formal description of the addition and subtraction operations can be given within the mathematical formulation of quantum mechanics. 
We define generalized subtraction and addition operators as 
\begin{eqnarray}
\hat{c}_k\equiv \sum^{\infty}_{n=1}k_n|n-1\rangle\langle n|,~~
\hat{c}^{\dag}_k =\sum^{\infty}_{n=1}k^{*}_n|n\rangle\langle n-1|,
\end{eqnarray}
where $n$ is the number of particles, and the addition operator $\hat{c}^{\dag}_k$ is derived by the adjoint of the subtraction operator.
The differences between quantum and classical systems are typically ascribed to the commutation relation of the addition and subtraction, which vanishes for classical particles. 
Using the relation (2.1), we can test the commutator,
 \begin{eqnarray}
 [\hat{c}_k,\hat{c}^{\dag}_k]=\sum^{\infty}_{n=1}(|k_n|^2-|k_{n-1}|^2)|n-1\rangle\langle n-1|,
 \end{eqnarray}
where $k_0\equiv0$. To make the right-hand side zero, we can put all the coefficients $k_n$ equal, i.e., $k_n=k_{n+1}$ ($n\geq 1$), such that 
we obtain the relation, $[\hat{c}_k,\hat{c}^{\dag}_k]=|k_1|^2|0\rangle\langle 0|$. Then the coefficient $k_1$ should be zero to make the commutator vanish.
However, the operator vanishes in this case, and one sees that the commutation relation cannot be made to disappear completely in this classical regime. 
The situation is very different in the quantum regime. For bosonic particles, the commutator is equal to the identity, $[\hat{c}_k,\hat{c}^{\dag}_k]=I$, with $|k_n|^2=n$. 
For fermionic particles, due to the Pauli exclusion principle, the operator is transformed into $\hat{c}_k\equiv k_1|0\rangle\langle 1|$ such that
the anticommutation relation becomes $\{\hat{c}_k,\hat{c}^{\dag}_k\}=|k_1|^2(|0\rangle\langle 0|+|1\rangle\langle 1|)$. 
Thus, the anticommutator is equal to the identity with $|k_1|=1$.

We have shown that although quantum formalism readily admits the description in terms of commutation relation, the generalization to the classical regime is not straightforward.
It can be understood by looking into the two terms $\hat{c}_k\hat{c}^{\dag}_k$ and $\hat{c}^{\dag}_k\hat{c}_k$ separately. Based on the relation (2.1), we obtain the following relations:
\begin{eqnarray}
&&\hat{c}_k\hat{c}^{\dag}_k=\sum^{\infty}_{n=1}|k_n|^2|n-1\rangle\langle n-1|,\\
&&\hat{c}^{\dag}_k\hat{c}_k=\sum^{\infty}_{n=1}|k_n|^2|n\rangle\langle n|.
\end{eqnarray}
For $|k_n|^2 =1$, the relations become $\hat{c}_k\hat{c}^{\dag}_k=I$, and $\hat{c}^{\dag}_k\hat{c}_k=I-|0\rangle\langle 0|$ respectively.
From this, one immediately sees that the commutation relation cannot approach zero because of the vacuum component in the addition-then-subtraction operation. 

Therefore, instead of the commutation relation, we propose to consider the addition-then-subtraction operation which reduces to the identity for classical particles.
For $m$ initial classical particles $|m\rangle_c$ ($m\geq 0$), we obtain the result $\hat{c}_k\hat{c}^{\dag}_k |m\rangle_c=|m\rangle_c$ with the relation (2.3) and $|k_n|^2=1$.
For $m$ bosonic particles $|m\rangle_b$ ($m\geq 0$),  $\hat{c}_k\hat{c}^{\dag}_k|m\rangle_b=(m+1)|m\rangle_b$ with $|k_n|^2=n$.
For $m$ fermionic particles $|m\rangle_f$ ($m=0$ or $1$),  $\hat{c}_k\hat{c}^{\dag}_k|0\rangle_f=|0\rangle_f$ and $\hat{c}_k\hat{c}^{\dag}_k|1\rangle_f=0$ with 
the Pauli exclusion principle and $|k_1|^2=1$. 
Therefore, by looking at the norm of the state $\hat{c}^{\dag}_k|m\rangle$, we can figure out whether a given particle is classical, bosonic, or femionic:
\begin{eqnarray}
_c\langle m|\hat{c}_k\hat{c}^{\dag}_k|m\rangle_c &=&1,~|k_n|^2=1, \nonumber\\
_b\langle m|\hat{c}_k\hat{c}^{\dag}_k|m\rangle_b &=&m+1,~|k_n|^2=n, \nonumber\\
_f\langle m|\hat{c}_k\hat{c}^{\dag}_k|m\rangle_f &=&0~or~1,~|k_1|^2=1.
\end{eqnarray}
The general relations for classical and quantum particles are shown in Table I,
where $\hat{c}_o$, $\hat{c}_b$, $\hat{c}_f$ are for $|k_n|^2=1$, $|k_n|^2=n$, $|k_1|^2=1$ respectively. 
Motivated by the above equations, we quantify the degree of indistinguishability
\begin{eqnarray}
I_d\equiv \langle m|\hat{c}_k\hat{c}^{\dag}_k|m\rangle,
\end{eqnarray}
where $m$ is the number of particles and classical, bosonic, and fermionic particles correspond to $I_d=1$, $I_d\geq 1$, and $0\leq I_d\leq 1$, respectively.

\begin{table}[]
\caption{Classical and quantum particles} \label{tab: title}
\begin{tabular}{|c|c|c|}
\hline Particle & (Anti) commutation  & \large{$\hat{c}_k\hat{c}^{\dag}_k$} \\
\hline  Classical  & $[\hat{c}_o,\hat{c}^{\dag}_o]=|0\rangle\langle 0|$ & $I$ \\
 Bosonic & $[\hat{c}_b,\hat{c}^{\dag}_b]=I$ & $\sum^{\infty}_{n=1}n|n-1\rangle\langle n-1|$ \\
 Fermionic & $\{ \hat{c}_f,\hat{c}^{\dag}_f\}=I$ & $|0\rangle\langle 0|$ \\
\hline
\end{tabular}
\end{table}

\section{Test of bosonic and fermionic properties via addition-then-subtraction operation}
In this section, we investigate the difference between bosonic and fermionic properties via the addition-then-subtraction operation and propose implementable schemes for bosons in  all-optical and atom-cavity systems.
The addition and subtraction operations are one-sided unitary, $\hat{c}_o\hat{c}^{\dag}_o=I$, in the classical limit, i.e., the addition operation is norm preserving for classical particles. For quantum particles, however, this is not the case. Thus, based on the norm preserving property, 
we can write down an analytic relation for bosons and fermions.

\subsection{Bosonic and fermionic properties}
The norm increases after the addition operation for bosonic particles, whereas it decreases for fermionic ones, except the initial vacuum state.   
It is convenient to write it in the following form: 
\begin{eqnarray}
_b\langle \psi |\hat{c}_b\hat{c}^{\dag}_b|\psi\rangle_b &\geq& _b\langle \psi|\psi\rangle_b, \\
_f\langle \psi |\hat{c}_f\hat{c}^{\dag}_f|\psi\rangle_f &\leq& _f\langle \psi|\psi\rangle_f, 
\end{eqnarray}
where $b$ $(f)$ denotes a boson (fermion), and the equality holds for the vacuum state.
We can test the inequalities with a density matrix, $\rho=(1-p)|0\rangle\langle 0|+p|1\rangle\langle1|$,
which can be considered in the case of bosons as well as fermions.
This density matrix can be generated with quantum scissors \cite{Pegg} and a thermal field, 
\begin{eqnarray}
\rho_{out}=_c\langle 0|_b\langle 1|\hat{B}_{bc}\hat{B}_{ab}\rho_{in}
\hat{B}^{\dag}_{ab}\hat{B}^{\dag}_{bc}|1\rangle_b|0\rangle_c,
\end{eqnarray}
where $\rho_{in}=|1\rangle_a\langle 1|\otimes |0\rangle_b\langle 0|\otimes \rho^c_{th}$, and
$\rho_{th}=\frac{1}{(1+\overline{n})}\sum^{\infty}_{n=0}(\frac{\overline{n}}{1+\overline{n}})^n|n\rangle\langle n|$ ($\overline{n}$ is the mean photon number).
The beam splitter $\hat{B}_{ab}$ ($\hat{B}_{bc}$) transforms the input modes as $\hat{a}^{\dag} \rightarrow t\hat{a}^{\dag}+r\hat{b}^{\dag}$ and 
$\hat{b}^{\dag} \rightarrow t\hat{b}^{\dag}-r\hat{a}^{\dag}$ [ $\hat{b}^{\dag}\rightarrow \frac{1}{\sqrt{2}}(\hat{b}^{\dag}-\hat{c}^{\dag})$ and
$\hat{c}^{\dag}\rightarrow \frac{1}{\sqrt{2}}(\hat{c}^{\dag}+\hat{b}^{\dag})$ ]. 
Coefficients $t$ and $r$ are the transmissivity and reflectivity of the beam splitter $\hat{B}_{ab}$, respectively.
 Thus the density matrix becomes 
$\rho_{out}=(1-p)|0\rangle\langle 0|+p|1\rangle\langle1|$ with $p=\frac{\frac{\overline{n}}{1+\overline{n}}|t|^2}{|r|^2+(\frac{\overline{n}}{1+\overline{n}})|t|^2}$.
When applying it to the inequalities (3.1) and (3.2), the left-hand side becomes $1+p$ for bosons and $1-p$ for fermions.
Note that this measure is different from the measure proposed by Kurzy\'nski et al. \cite{Pawel}, which is optimized for the density matrix $\rho$ at $p=1/3$.
On the other hand, our measure is optimized for $\rho$ at $p=1$.

For bosons, we can test other quantum states, such as a coherent state $|\alpha\rangle$ or a thermal state $\rho_{th}$.
The left-hand side of the inequality (3.1) becomes $1+|\alpha|^2$ and $1+\overline{n}$ for coherent and thermal states, respectively.
We can see that the degree of indistinguishability for the states increases with the mean photon number of each state.
Furthermore, we can consider higher-order relations as follows:
\begin{eqnarray}
_b\langle \psi |\hat{c}^n_b\hat{c}^{\dag n}_b|\psi\rangle_b &\geq& _b\langle \psi|\psi\rangle_b,
\end{eqnarray}
where $n$ is a positive integer.
Since the operator $\hat{c}_b$ is equal to the photon annihilation operator $\hat{a}$, the left-hand side of the inequality (3.4) becomes the expectation value of the antinormal ordering operator,
$\langle \hat{a}^n\hat{a}^{\dag n}\rangle=\int^{\infty}_{-\infty}d^2\alpha Q(\alpha) |\alpha |^{2n}$, $Q(\alpha)=\frac{1}{\pi}\langle \alpha |\rho |\alpha\rangle$,
where $Q(\alpha)$ is positive semidefinite \cite{Zubairy}.
For coherent and thermal states, the l.h.s. of the inequality (3.4) becomes $n!L_n(-|\alpha|^2)$ and $n!(1+\overline{n})^n$, respectively, where $L_n(-|\alpha|^2)$ is the Laguerre polynomial.
For a given mean photon number, their indistinguishability increases much more rapidly with increasing $n$.

\subsection{Implementable schemes}
As we have discussed earlier, we do not consider the operation $\hat{a}^{\dag}\hat{a}$. Thus, for consistency, the expectation value  $\langle \psi |\hat{a}\hat{a}^{\dag}|\psi\rangle$ should be measured without resorting to the commutation relation. 
For this purpose, we consider coupling a probe to a bosonic system $|\psi\rangle$. This indirect measurement requires only a two-level probe, whereas
a direct measurement requires a multilevel probe. 
Here, we show that it is possible to obtain the expectation value $\langle \psi | \hat{a}\hat{a}^{\dag}| \psi \rangle$ by counting the photon number of the idler mode in a nondegenerate parametric amplifier (NDPA) with small coupling strength. 
Likewise, the expectation value is also obtained by detecting the atomic state of the atom-cavity field system at $gt\sqrt{n+1}\ll 1$
if the atom is initially prepared in the upper state $|e\rangle$.

In an optical system, the norm of bosonic particles via addition operation can be obtained with the help of an NDPA
with small coupling strength $s \ll 1$. An arbitrary state $|\psi\rangle$ is injected to a single mode of the NDPA with the idler mode in the vacuum state,
\begin{eqnarray}
\exp(-s\hat{a}^{\dag}\hat{b}^{\dag}+s\hat{a}\hat{b})|\psi\rangle_a |0\rangle_b
\approx (1-s\hat{a}^{\dag}\hat{b}^{\dag})|\psi\rangle_a |0\rangle_b.
\end{eqnarray}
In this case, there are only two possibilities on the number of photons in mode b, as verified in a recent experiment \cite{Parigi}.
Then, counting the number of photons in mode b, we get the probability of each outcome, $P_b(0)$ and $P_b(1)$, as follows:
\begin{eqnarray}
&&P_b(0)=\frac{\langle \psi |\psi\rangle}{\langle \psi |\psi\rangle +s^2\langle \psi |\hat{a}\hat{a}^{\dag} |\psi\rangle},\\
&&P_b(1)=\frac{s^2\langle \psi |\hat{a}\hat{a}^{\dag} |\psi\rangle}{\langle \psi |\psi\rangle +s^2\langle \psi |\hat{a}\hat{a}^{\dag} |\psi\rangle}.
\end{eqnarray}
Given a coupling strength of the NDPA, we can obtain the norm of bosonic particles after the addition operation by counting the number of photons in mode b,
\begin{eqnarray}
\langle \psi |\hat{a}\hat{a}^{\dag} |\psi\rangle=\frac{P_b(1)}{s^2P_b(0)}=\frac{N_b(1)}{s^2N_b(0)},
\end{eqnarray}
where $P_b(0)=\frac{N_b(0)}{N_b(0)+N_b(1)}$ and $P_b(1)=\frac{N_b(1)}{N_b(0)+N_b(1)}$. 
We assume that the norm of the initial state is equal to 1, i.e., $\langle \psi |\psi\rangle=1$.
$N_b(0)$ and $N_b(1)$ are the number of counts that zero and one photon are detected in mode b, respectively.
At $s=0.01$ \cite{Parigi}, the probability of detecting more than one photon is negligible and one would see no detection event except for the few times when one photon is detected in the idler mode b.  By considering an on-off detector efficiency $\eta$, Eq. (3.8) becomes $\frac{\eta N_b(1)}{s^2N_b(0)}$.

In an atom-cavity system, we can obtain the norm by interacting a two-level atom with a single-mode cavity field at resonance.
The time evolution of the atom-field system is 
$|\psi (t)\rangle = e^{-i\hat{H}t/\hbar}|\psi (t=0)\rangle$, where $\hat{H}=\hbar g(\hat{a}^{\dag}\hat{\sigma}_{-}+\hat{\sigma}_{+}\hat{a})$,
$\hat{\sigma}_{-}=|g\rangle\langle e|$, and $\hat{\sigma}_{+}=|e\rangle\langle g|$  \cite{Zubairy}.
$g$ is the atom-field coupling constant,  $\hat{\sigma}_{-}$ and  $\hat{\sigma}_{+}$ are atomic ladder operators with $|e\rangle$ and $|g\rangle$ denoting
the upper and lower states of the atom, and $\hat{a}^{\dag}$ and $\hat{a}$ are photon creation and annihilation operators.
Initially, the atom is prepared in $|e\rangle$ and the cavity field is in an arbitrary pure state $|\psi\rangle=\sum_{n=0}c_n(0)|n\rangle$. 
Then, the atom-cavity system evolves as \cite{Jiyong}
\begin{eqnarray}
|\phi (t)\rangle&=&|e\rangle \otimes \sum_{n=0}c_n(0)\cos(gt\sqrt{n+1})|n\rangle \nonumber\\
&&-i|g\rangle \otimes \sum_{n=0}c_n(0) \frac{\sin(gt\sqrt{n+1})}{\sqrt{n+1}}\hat{a}^{\dag}|n\rangle.
\end{eqnarray}
At $gt\sqrt{n+1}\ll 1$, 
\begin{eqnarray}
|\phi (t)\rangle &\approx& |e\rangle \otimes \sum_{n=0}c_n(0)|n\rangle
-igt |g\rangle \otimes \sum_{n=0}c_n(0) \hat{a}^{\dag}|n\rangle \nonumber\\
&=& |e\rangle \otimes |\psi\rangle -igt|g\rangle \otimes \hat{a}^{\dag}|\psi\rangle.
\end{eqnarray}
Equation (3.10) can be approximately satisfied with small coupling constant ($g$), and short interaction time ($t$) for a given maximum photon number. 
Then, detecting the atomic state, we obtain the norm
\begin{eqnarray}
\label{eq3.11}
\langle \psi |\hat{a}\hat{a}^{\dag}|\psi\rangle =\frac{N_g}{|gt|^2 N_e},
\end{eqnarray}
where $N_g$ ($N_e$) is the counting number of the atom in the lower (upper) state.
For a coherent state, the condition equivalent to $gt\sqrt{n+1}\ll 1$ is achieved by keeping the amplitude $|\alpha|$ of the coherent state small and keeping the interaction time short, which can be controlled by the speed of an atom injected into a cavity \cite{Haroche}.
The atoms passing through the cavity field will be detected in the upper state most of the time with only a small number of them detected in the lower state.
The probability of detecting the output atom in the lower state increases as $1+\langle \hat{a}^{\dag}\hat{a}\rangle$ as shown in Eq.~(\ref{eq3.11}).
When considering two identical field ionization detectors \cite{Haroche1}, Eq.~(3.11) remains unchanged even with a finite detection efficiency.

\section{Conclusion}
We have shown how to distinguish quantum particles from classical particles via an addition-then-subtraction operation.
With a generalized operator description, we have demonstrated the difference between the norm of the particles $\langle \psi|\hat{c}_k\hat{c}^{\dag}_k|\psi\rangle$. For one-particle states,
the norm becomes $1$, $2$, and $0$ for classical, bosonic, and fermionic particles, respectively. 
Based on our definition (2.6), the degree of indistinguishability for bosonic particles increases with the number of particles, and the one for fermionic particles exhibits a range $0\leq I_d\leq 1$. For classical particles, the degree of indistinguishability is fixed as $1$. 
We have applied the addition-then-subtraction operation to quantify bosonic and fermionic properties, showing in particular that the indistinguishability increases with the mean photon number.
Furthermore, we have shown that one can evaluate the deviation from classical particles, with an indirect measurement requiring only a two-level probe,
by measuring an idler mode of a nondegenerate parametric amplifier with small coupling strength or
by detecting an atomic state of a weakly interacting atom-cavity field system. 

The ability to determine the degree of indistinguishability of bosonic or fermionic systems is important if one wants to use these systems in quantum computation and quantum information processing. 
For instance, boson sampling \cite{bosonsampling} relies on the indistinguishability of bosons to compute the permanent of a matrix with an exponential speedup with respect to classical computers.

\begin{acknowledgments}
S.Y.L. thanks P. Kurzy\'nski for helpful discussions.
This work was supported by the National Research Foundation and Ministry of Education in Singapore.
\end{acknowledgments}

\end{document}